\documentclass[12pt,preprint]{aastex}
\slugcomment{{\em}}

\def\aa{{\it Astron. Astrophys.} \,}
\def\apj{{\it Ap. J.} \,}
\def\apjs{{\it Ap. J. Supp.} \,}

\def\apjl{{\it Ap. J. Lett.} \,}
\def\mn{{\it MNRAS} \,}
\def\aj{{\it Astron. J.} \,}
\def\ph{{\it astro-ph/}}

\def\aa{{\it Astron. Astrophys.} \,}
\def\apj{{\it ApJ.} \,}
\def\apjs{{\it Ap. J. Supp.} \,}

\def\apjl{{\it Ap. J. Lett.} \,}
\def\mn{{\it MNRAS} \,}
\def\aj{{\it Astron. J.} \,}

\def\pasj{{\it PASJ} \,}
\def\ph{{\it astro-ph/}}

\def\pasj{{\it Publ. Astron. Soc. Japan \,}}

\def\eg{{\it e.g.\,}}

\begin{document}
%\title{A New Approach For Determining Clusters Properties}
\title{A New Approach for Simulating Galaxy Cluster Properties}

\author{Y. Arieli$^{1,2}$,  Y. Rephaeli$^{1,2}$ \& M. L. Norman$^2$}
 
\affil{$^1$ School of Physics and Astronomy, Tel Aviv University, 
Tel Aviv, 69978, Israel \\
$^2$ Center for Astrophysics and Space Sciences, University of 
California, San Diego, La Jolla, CA 92093-0424 \\ }

\begin{abstract}

We describe a subgrid model for including galaxies into hydrodynamical 
cosmological simulations of galaxy cluster evolution. Each galaxy 
construct-- or {\em galcon}-- is modeled as a physically extended object 
within which star formation, galactic winds, and ram pressure stripping 
of gas are modeled analytically. Galcons are initialized at high redshift 
($z \sim 3$) after galaxy dark matter halos have formed but before the 
cluster has virialized. Each galcon moves self-consistently within the 
evolving cluster potential and injects mass, metals, and energy into 
intracluster (IC) gas through a well-resolved spherical interface layer. 
We have implemented galcons into the {\em Enzo} adaptive mesh refinement 
code and carried out a simulation of cluster formation in a $\Lambda$CDM 
universe. With our approach, we are able to economically follow the 
impact of a large number of galaxies on IC gas. We compare the results of the galcon 
simulation with a second, more standard simulation where star formation 
and feedback are treated using a popular heuristic prescription. One 
advantage of the galcon approach is explicit control over the star 
formation history of cluster galaxies. Using a galactic SFR derived 
from the cosmic star formation density, we find the galcon simulation 
produces a lower stellar fraction, a larger gas core radius, a more 
isothermal temperature profile, and a flatter metallicity gradient 
than the standard simulation, in better agreement with observations. 

\end{abstract}


\section{Introduction}
\label{sec:intro}

Current hydrodynamic cosmological simulations of galaxy clusters show 
an appreciable level of inconsistency with results from high-precision 
optical and X-ray observations. Discrepancies are particularly apparent 
in intracluster (IC) gas properties 
%%(e.g., Tornatore et al. 2003, Borgani et al. 2003)
%(for recent review references see Borgani et al. 2008)
(e.g., Tornatore et al. 2003, Kay et al. 2007, Tornatore et al. 2007; 
for a recent review, see Borgani et al. 2008) 
- spatial distributions of density, temperature, and metallicity, but 
also in the stellar component for which simulations usually over-predict 
the stellar mass fraction while underpredicting the total number of 
galaxies (Nagamine et al. 2004 and references therein.). Physical 
processes, such as galactic winds, ram-pressure stripping, mergers of 
subclusters, energetic particle heating, and gravitational drag, affect 
the dynamical and thermal state of IC gas. Several attempts to implement 
some of these phenomena have been made (e.g., Kapferer et al. 2006, 
Domainko et al. 2006, Bruggen \& Ruszkowski 2005, Sijacki \& Springel 
2006, Cora 2006, Kapferer et al. 2007), with some success in 
reconstructing IC gas properties. However, different combinations of 
these processes and their specific implementation in numerical codes 
generally result in quite different gas properties. 

Because modeling star formation (SF) self-consistently requires 
prohibitively high level of spatial resolution, most current simulations 
use a SF prescription that follows the formation of collisionless star 
`particles' which feedback mass and energy to IC gas (Cen \& 
Ostriker 1992; Nagai \& Kravtsov 2005).
This approach overestimates the SF rate (SFR) at low z (Nagamine 
et al. 2004), which leads to a higher than expected star to gas mass 
ratio. In addition, feedback from the star particles is unresolved, 
leading to unrealistically low levels of gas (including metals) and 
energy transfer from galaxies into IC gas, and consequently insufficient 
suppression of cooling and gas overdensity in cluster cores. This 
unsatisfactory state motivates our attempt to develop a new method 
that partly overcomes current numerical limitations.

In this {\em Letter} we briefly describe a new approach (Section 2) for 
including galaxies in hydrodynamical cosmological simulations of 
cluster evolution which provides improved control over the relevant 
physical processes. Galaxies which are otherwise under-resolved (or 
absent!) are replaced with a physically-extended galaxy subgrid model 
which we refer to as a {\em galcon} within which SF, galactic winds, and 
ram pressure stripping of gas are modeled analytically. Galcons are 
initialized at high redshift after galaxy dark matter (DM) halos have 
formed but before the cluster has virialized. Mass, metals, and energy 
are injected from galcons into IC gas. In Section 3 we compare the 
results of our galcon simulation with a standard simulation using a 
popular star formation and feedback recipe, and summarize our main 
conclusions in Section 4.

\section{Simulation and Modeling Procedures}
\label{model}

Cluster evolution is followed using Enzo, a powerful adaptive mesh 
refinement (AMR) cosmological hydrodynamical code (Bryan \& Norman 1997). 
A high resolution cosmological simulation with baryons is performed 
starting at an initial redshift $z_i \simeq 60$; the evolution is 
stopped at $z_r=3-5$ (the 'replacement' redshift), where we know from 
observations the SFR peaked and early galaxies were already highly 
developed. At this time galactic halos with total mass $10^9-10^{12} 
M_\odot$ within the Lagrangian volume of the cluster are identified 
by a halo finding 
%
%technique 
technique. Since the dyanmics of DM halos is followed, they are allowed 
to merge, but galcons in merged halos are still identified as separate 
systems. Also, no new galcons are created at $z<3$, but this hardly 
matters since high-mass galaxies already formed by $z \sim 3$.
%
%(Eisenstein \& Hut 1998), 
and the baryon density profiles are fit by $\beta$ models. Galcons with 
these analytic density profiles are inserted into the centers of each 
halo, and assigned the halo velocity. Each galcon's central density and 
outer radius are determined from the fit and the baryonic mass within the 
halo virial radius. An equal amount of baryons is removed from the 
simulated density field. Note that the total mass density field, which 
is the sum of DM, baryonic gas, and stars is not affected by this 
replacement. Thus, an unphysical instantaneous change in the simulated 
density distribution does not occur. 

Both stellar and gaseous components are included in galcons, and since 
%
%their spatial distributions are expected to be initially very similar
%it is reasonable to take for both 
stars form in the same high gas density interstellar (IS) central regions 
that contain most of the gas and can be assumed to have initially 
roughly similar spatial distributions, it is reasonable to approximate 
both by  
the same $\beta$-profile parameters, but with different central densities. 
The mean initial baryonic mass density in galaxies can be determined by 
multiplying the mass density of halos from the Press \& Schechter (PS, 
1974) mass function by the universal baryonic density parameter $\Omega_b$. 
The stellar mass density is calculated by integrating the cosmic SFR 
density (to be specified below) over the interval $[z_{i}, z_{r}]$. 

Having initialized galcons, we follow their motion dynamically using 
Enzo's N-body machinery and follow the mass and energy ejection processes 
that enrich and heat up IC gas - galactic winds and ram pressure 
stripping -- through simple analytic models. Galactic winds reduce the 
total stellar mass while ram pressure stripping continuously reduces the 
galcon outer gas radius (as quantified below). Since galactic winds are 
SN driven, their elemental abundances are higher than in IS gas by a 
factor of $\sim 3$. We follow the enrichment by both processes, separately 
and jointly.

Observations of galactic winds (\eg, Heckman 2003) provide direct 
evidence for the relation between the mass and energy ejection 
rates and the SFR, $\dot{M}_\ast$,
\begin{eqnarray}
\label{eq:wind_ejecta}
\dot{E}_{w}=e_{w} \dot{M}_\ast c^2\; \nonumber \\
\dot{M}_{w}=\beta_{w} \dot{M}_\ast \;,
\end{eqnarray}
where the parameters $e_{w}$ and $\beta_{w}$ are energy and mass 
ejection efficiencies, respectively, and $c$ is the speed of light. 
The ejection efficiencies cannot be directly predicted from simple 
considerations, but they can be roughly estimated from observations 
(e.g., Pettini et al 2001; Heckman et al. 2001), from which typical 
values of $e_{w}=5\times 10^{-6}$ and $\beta_{w}=0.25$ are adopted 
(e.g., Cen \& Ostriker 1993, Leitherer et al. 1992).

An advantage of the galcon approach is analytic control over the SF 
history of each galaxy which we take from observations. The galactic 
SF history is determined indirectly from the closely related stellar 
mass density, which can be estimated by fitting multiband photometric 
observations to simulated galaxy spectra generated by a population 
synthesis model (e.g., Brinchmann \& Ellis 2000, Cohen 2002, Glazebrook 
et al. 2004). Ideally, we would use the SF history of observed cluster 
galaxies for our galcons. As this information is not yet available to 
us, we instead use the observed cosmic SF density to illustrate our 
method. It has been suggested (Nagamine et al. 2006) that the cosmic 
SF density can be expressed as the sum of two exponential terms based 
on the characteristic SF times of disk and spheroid galaxies, where the 
scaling between the two components is determined by the spheroid to disk 
stellar mass ratio. The mean SFR in a galaxy can then be roughly 
estimated from the universal SFR rate by simply taking the ratio of the 
mass density of galaxies - determined from the PS mass function - to the 
observationally deduced universal SFR per unit volume. The 
two-parameter fit 
$s(t) = 1.58 \times 10^{-2} t^{\alpha} exp(-\gamma t) \,$ 
$Gyr^{-1}$, with (cosmological time) $t$ in Gyr, $\alpha = -0.70$, 
and $\gamma = 0.07$, represents well the SFR density. The total SFR in 
a galaxy at a given time, which is used to set the energy and mass 
ejection by galactic winds, can then be calculated by multiplying the 
above SFR density by the total mass of the galaxy.

SN ejecta from the galactic disk quickly interact through shocks and mix 
with the surrounding IS gas. Thus, the wind contains a blend of
metals from stars and IS medium. However, the fraction of gas in the 
ejecta cannot accurately be determined. We assume that most of the wind 
ejecta come from the stellar component whose metallicity is approximately 
solar; accordingly, we subtract the energy and mass carried by the wind 
from the galcon stellar content, and increase the metallicity of IC gas 
by an amount which is proportional to galcon mass ejecta. The transfer of 
mass and energy to IC gas is implemented by isotropically distributing 
the ejecta over a thin shell - typically a few kpc - surrounding the 
outer galcon radius. Also, since the level of SN activity is roughly 
linearly proportional to the local star density, the wind does not modify 
the spatial stellar profile. Thus, we only adjust the central density of 
the stellar component to reflect the loss of stellar material.

Ram pressure stripping is implemented by determining (at any given time) 
the stripping radius, where the local IC gas pressure is equal to the 
local galactic IS pressure, simply assuming that all IS gas outside this 
radius is stripped on a relatively short dynamical timescale. We 
generalize the analytic Gunn \& Gott (1972) stripping condition by 
including the (generally dominant) contribution of DM to the galactic 
gravitational force (which was ignored in some previous works). 
Observational evidence supports the expectation that stripping truncates 
the gaseous disk but does not modify the gas profile; neither does it 
appreciably affect the dynamics of the stellar and DM components of the 
galaxy (Kenney \& Koopmann 1999; Kenney, Van Gorkom \& Vollmer 2004). 
Thus, the outer radius of the galcon gas component is reduced to the 
stripping radius without modifying the central density or the scale 
radius of its profile. 

\section{Results}
\label{results}

\begin{figure}
\epsscale{0.9}
\plotone{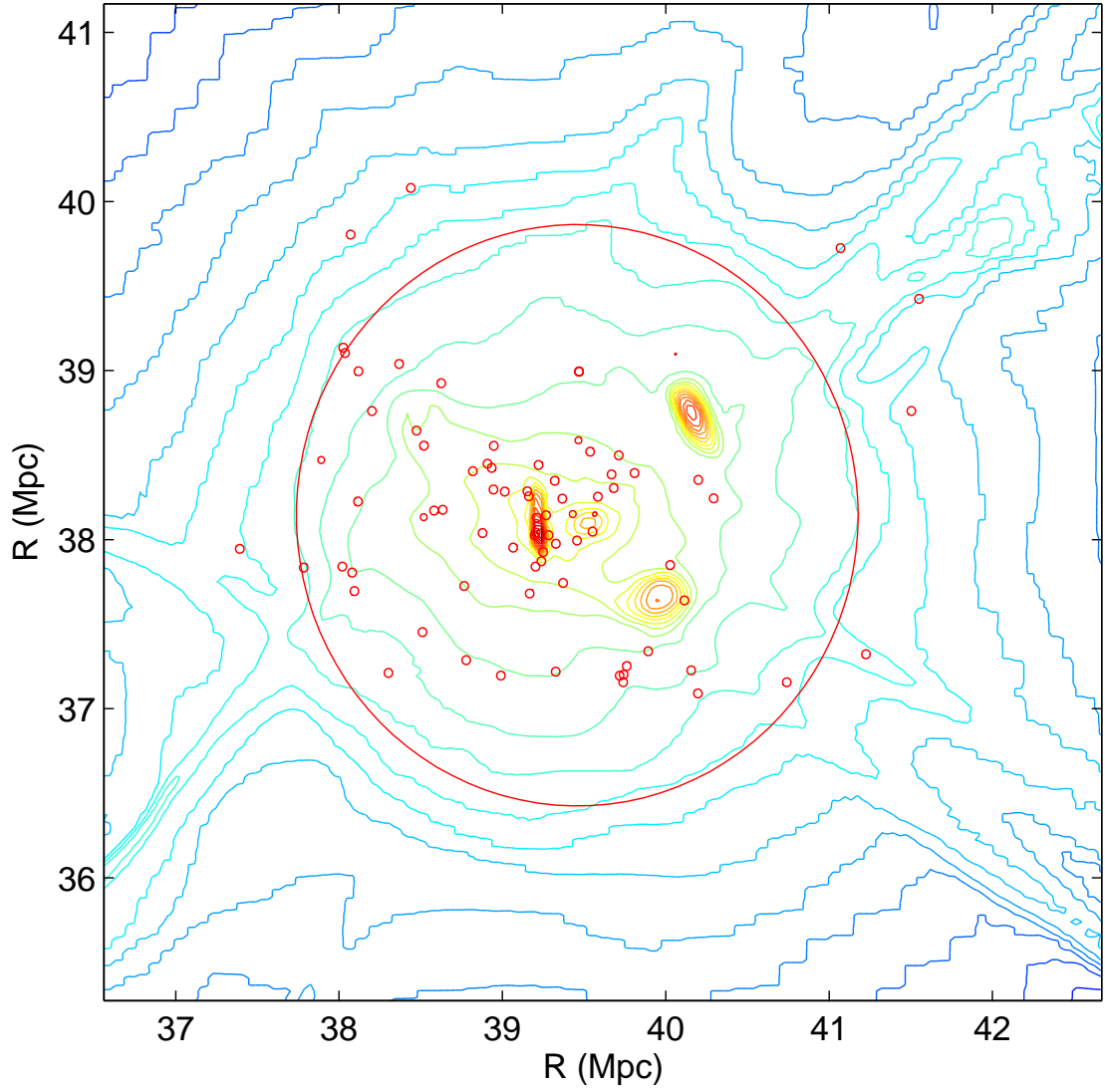}\label{fig:positions}
\figcaption{Positions of galcons at z=0 are shown over a map of 
projected gas density iso-density contours. The red circle indicates 
the virial radius, $R_{v}=1.71 \; Mpc$.}
\end{figure}

In order to quantify the improvements in the description of the evolution 
of IC gas in our new code, we have performed two high resolution runs 
with AMR and radiative cooling (Sutherland \& Dopita 1993). The first - 
hereafter the galcon run (GR) - included the additional physical 
processes and galcons, whereas the second - the comparison run (CR) - 
included the SF and feedback recipe of Cen \& Ostriker (1992). The setup 
of both runs includes a root grid of $128^3$ cells which covers a 
comoving volume of 54 $Mpc^3$ with two nested inner grids. The most 
refined grid covers a comoving volume of 27 Mpc$^3$ divided into 
$128^3$ cells and can be further refined by up to 5 levels, with a 
maximum $\sim$9 kpc resolution.

Both runs were initialized at $z=60$ assuming a $\Lambda$CDM 
model with $\Omega_m=0.27$, $\Omega_\Lambda=0.73$, $\sigma_8=
0.9$, and $h=0.71$ ($H_0$ in units of 100 $km\; s^{-1}\; 
Mpc^{-1}$). The CR was evolved continuously to $z=0$. The GR was 
stopped at $z=3$, and a halo-finding algorithm (Eisenstein \& Hut 
1998) was used to locate 89 galactic halos with mass in the 
range $10^9-10^{12}\; M_\odot$ within a volume which eventually 
collapsed to form the cluster. The baryonic content of these halos 
was analyzed and replaced by galcons, as described in section 
\ref{model}. The simulation was then evolved to $z=0$ with the 
additional physical processes and galcons. Both simulations included 
radiative cooling over the entire ($z \geq 0$) evolution. A fuller 
description of the simulation and the models will be given in a 
forthcoming paper (Arieli, Rephaeli \& Norman, in preparation). 

Relatively rich $\sim 5.4 \times 10^{14} \; M_\odot$ clusters 
with similar global properties were generated in both runs. Near equality 
in the global properties of the simulated clusters is expected since 
(identically treated) DM dynamics govern cluster formation and evolution. 
However, the composition of the cluster and the properties of the 
baryonic component, particularly in the core, are substantially 
different. A significant difference is seen in the number of cluster 
%galaxies. We 
galaxies; we 
identified a final number of 81 galaxies within the virial radius of 
the GR cluster (Fig. 1), whereas only 6 galaxies were identified in the 
CR cluster. 
Thus, 
The drastically lower number of identified galaxies in the CR cluster 
stems from inadequate force resolution which results in the unphysical 
merging of galaxy DM halos (the ``overmerging problem"; Moore et al. 
1996, Klypin et al. 1999.) Klypin et al. 1999 argue that a {\em proper} 
force resolution of $\leq 2 h^{-1}$ kpc and mass resolution 
$\leq 10^9 h^{-1} M_{\odot}$ is required for galaxy mass halos to 
survive in the dense cluster core. While our Enzo simulation meets the 
mass resolution requirement, its {\em comoving} force resolution is 
7.8 $h^{-1}$ kpc -- less than required. However, by replacing the 
baryon content of galaxies with galcons at z=3, where the proper force 
resolution is four times better, we ``lock in" their mass distribution 
in a resolution-independent way. During the N-body dynamics phase of the 
calculation, the galcon's extended mass distribution is deposited to the 
mesh where it helps anchor the DM halo despite less than optimal force 
resolution. 

Our galaxies remain intact as they pass through the cluster core as they 
are ``indestructable." This has important consequences for IC gas 
properties, as we discuss next.

\begin{figure}
\epsscale{1.1}
\label{fig:four_panels}
\plottwo{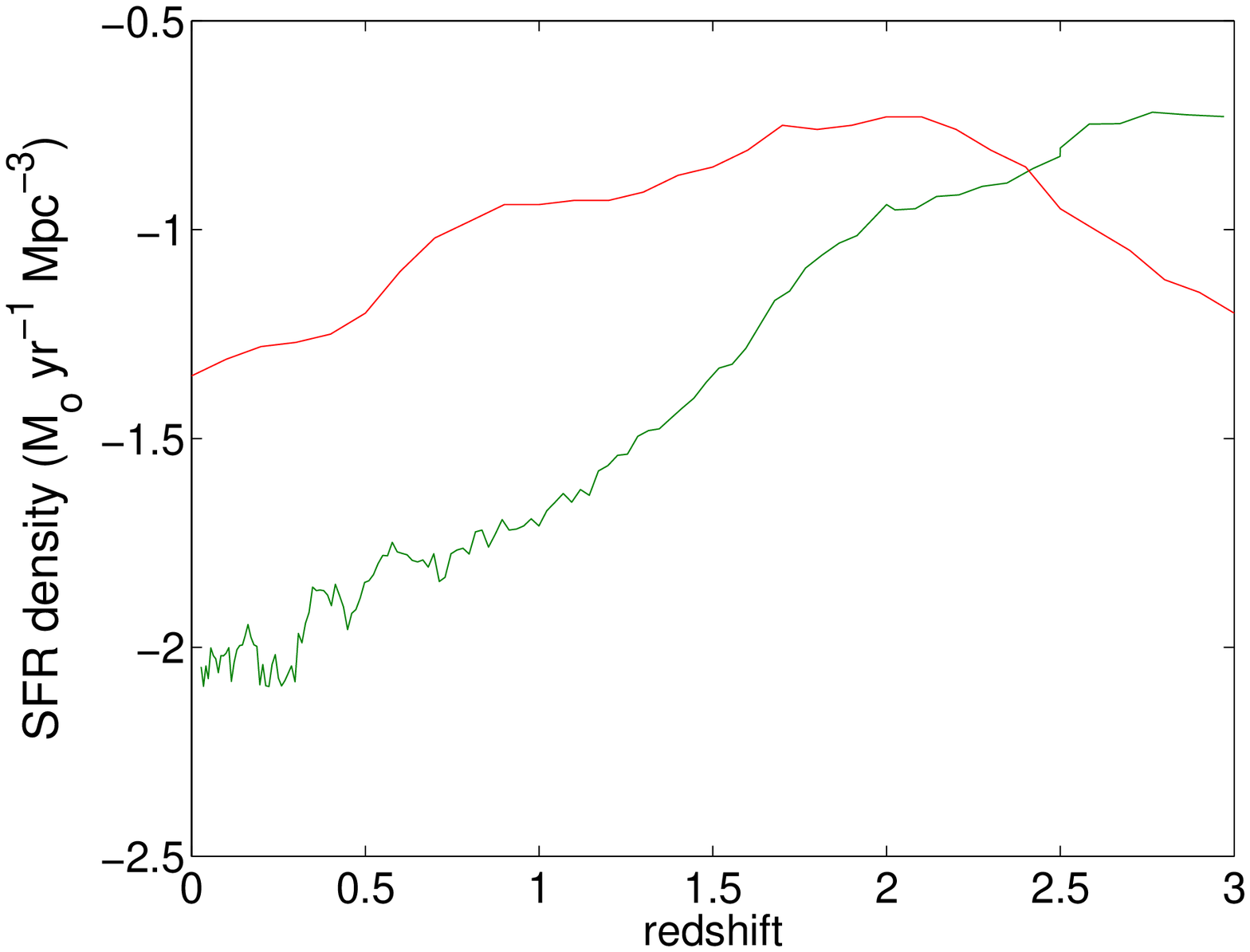}{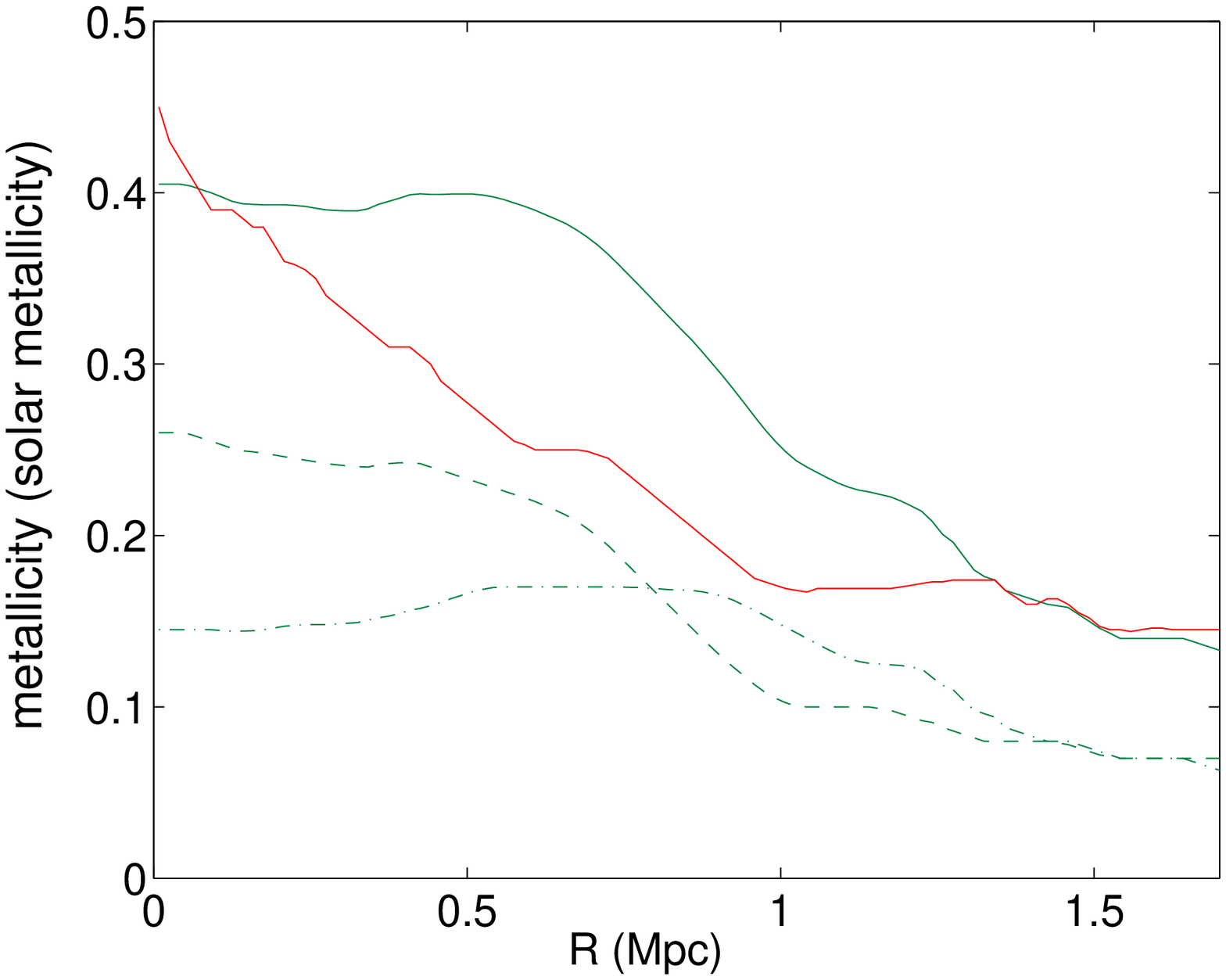}
\plottwo{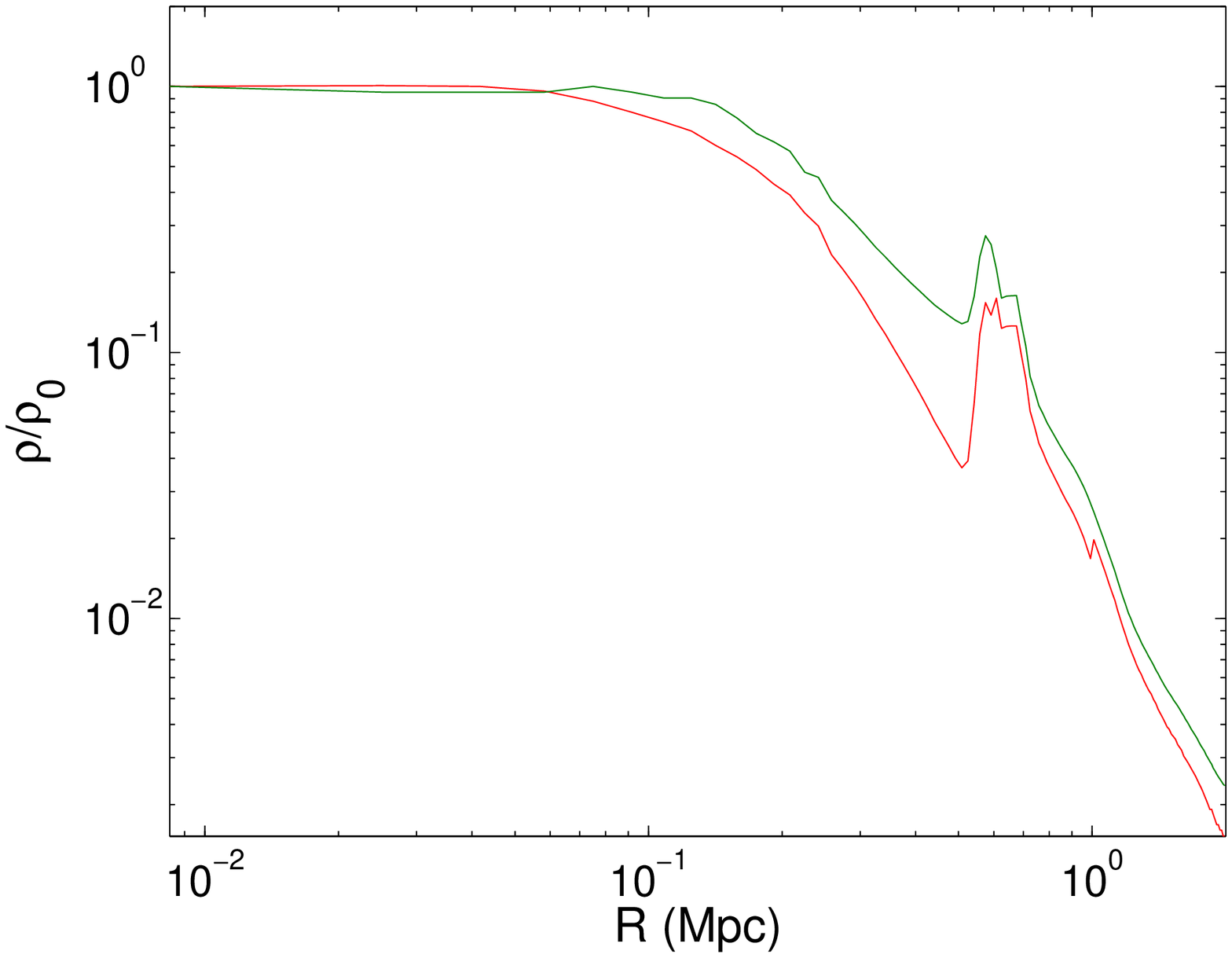}{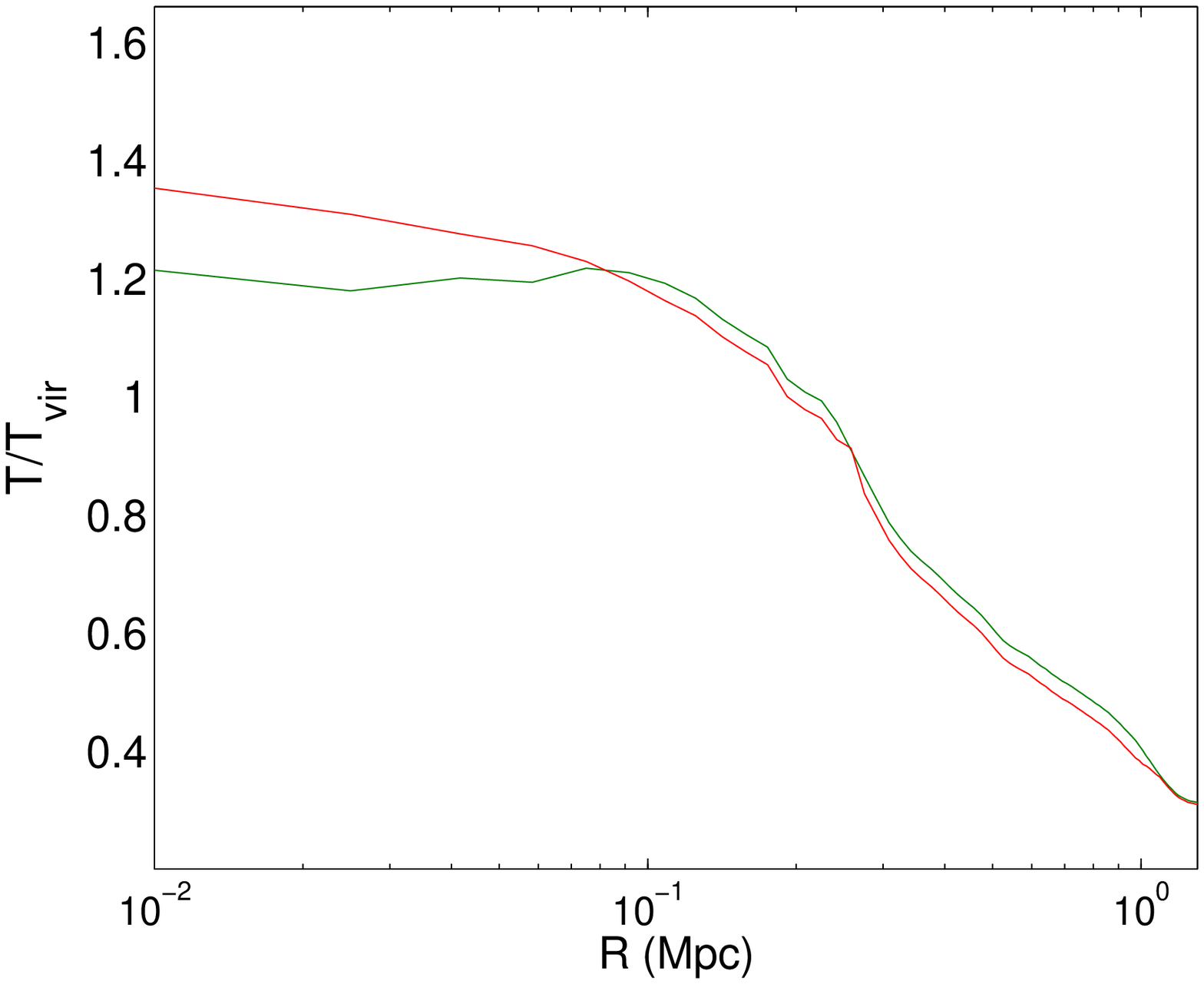}
\figcaption{The SFR density (top left) and the profiles of IC gas 
matallicity (top right), density (bottom left), and temperature 
(bottom right) in the simulated GR (green) and CR (red) clusters 
are shown in the four panels. Also shown are the separate contributions 
to the metallicity from wind ejecta (dashed-dotted line) and from ram 
pressure stripping (dashed line) in the GR cluster.} 
\end{figure}

\subsection{Star formation and heating}
\label{sec:star-formation}

SF history is a central driver of gas feedback processes. It is therefore 
very important to verify that our GR simulation produces the required 
heating to overcome overcooling, and results in a reasonable stellar
mass fraction. As stated before, the simulation does not directly 
input the observational cosmic SFR to each galaxy. Rather, it only 
incorporates the functional behavior of the SFR, with its actual 
level and history determined based on the galcon mass and redshift. 
The SFR density in the GR and CR clusters is shown in the top left 
panel of Fig. 2. The cosmic SFR density in 
the GR cluster is consistent with observations; the agreement is 
good at high $z$, whereas below $z \sim$0.5 the SFR density is at 
the lower end of the observationally deduced range. This could 
possibly be due to the lack of galcon mergers, events during which 
the SFR is enhanced. Analysis of the simulation does show that at $z=0$ 
%some $15\%$ of the 
a small number of 
galcons are sufficiently close to merge. However, mergers and the 
associated boost of SF are not implemented in the current version of 
the simulation; their impact will be explored in planned subsequent 
work. The star to gas mass ratio is $\sim$9.5\%, in good agreement 
with the observationally deduced value (e.g., Balogh et al. 2001, Wu 
\& Xue 2002).

\subsection{Gas metallicity, density and temperature}
\label{cluster_properties}

Metallicity in the GR cluster (top right panel of Fig. 
2) is due to enrichment by both winds and 
ram pressure stripping, with the former process being more effective 
at higher redshifts, since it is driven by shocks from SN that are 
then more prevalent. At these early periods of cluster evolution a 
higher fraction of galaxies are outside the cluster core, where metals 
are preferentially deposited. As the cluster evolves galaxies are more 
centrally distributed, so metals are more effectively spread in the 
central region. This results in an approximately constant metallicity 
across the cluster. On the other hand, because gas stripping depends on 
the local IC gas density, which builds up as the cluster evolves, the 
contribution to the metallicity is larger at lower $z$, and is more 
concentrated in the high density core, resulting in a substantial 
metallicity gradient.

The total metallicity in the GR cluster is roughly constant out to 
$\sim$700 kpc; it decreases at larger radii. Its mean value across 
the cluster is 0.32$Z_{\odot}$, within the observationally determined 
range, $(0.3-0.4)Z_{\odot}$. The mean metallicity in the CR cluster 
is 0.25$Z_{\odot}$, somewhat lower than typical. Moreover, its steep 
decline already in the central region is also at odds with observations, 
which show a nearly constant metallicity in the central few hundred kpc 
%(Hayakawa et al. 2006, Pratt et al. 2006).
(Hayakawa et al. 2006, Pratt et al. 2006), with the exception of a small 
galactic-size region at the cluster center where the metallicity is higher 
(Snowden et al. 2008). A much shallower gradient is observed in cooling 
flow clusters (e.g., De Grandi et al. 2004), but our simulated clusters 
have no cooling flows. We conclude that in our GR simulation - which 
includes galactic winds and ram pressure stripping - both the level of 
metallicity and its spatial profile are consistent with observations, 
whereas neither property is well reproduced in the CR cluster.

The density profiles (bottom left panel of Fig. 2) 
are similar at large radii, including a steep hump at $r \sim$600 kpc, 
indicating the location of a very massive clump. The profiles flatten 
towards the center, but the GR cluster has a substantially larger core 
of $\sim$180 kpc compared to a relatively small core of $50$ kpc in 
the CR cluster. The shape of the density profile in the central region 
is mainly determined by the fraction of IC gas that cooled down and 
converted into stars. The absence of sufficient feedback in the CR 
cluster results in too much cool gas in the inner core. Excessive 
cooling results in a small core as well as an unrealistically high 
number of stars, while the GR cluster includes stronger and more 
efficiently spread feedback, resulting in suppression of overcooling 
and in a larger core.

In a non cooling-flow cluster the core is isothermal and the temperature 
profile declines rapidly with radius outside the core (e.g., De Grandi 
\& Molendi 2002). This decrease is indeed seen in both clusters at 
radii larger than $\sim$200 kpc (bottom right panel of Fig. 
2). However, only the GR cluster has a flat 
isothermal core, while in the CR cluster the temperature continues a 
moderate rise towards 
the center. In the CR simulation a local SF prescription is used; this 
leads to formation of star groups which are at locations where the gas 
is dense and cold. Feedback heating from these star groups remains 
localized to their immediate inner core region, too concentrated to heat 
the outer core. This is particularly apparent at lower $z$. In contrast, 
in our GR simulation the implementation of winds out of galcons 
effectively spreads out the heating over a much larger volume, 
resulting in a large isothermal core. 

\section{Conclusion}

The combinination of our galaxy constructs and new semi-analytic 
modeling of the relevant physical processes yields a powerful tool that is 
capable of reproducing the basic properties of clusters. Our new approach 
successfully describes SF and the basic properties of IC gas, including 
its metallicity and energy feedback. The ever improving observational 
data motivate further development of the code and inclusion of additional 
physical processes previously unaccounted for, such as AGN feedback. We 
plan to improve the description of galactic mergers, and intend to 
implement an improved algorithm for replacing galactic halos with new 
galcons as the cluster evolves, instead of performing this replacement 
only at an initial redshift as has been done in the simulations reported 
here. Ongoing work on this project will hopefully lead to a much better 
understanding of the intrinsic properties of both DM and baryons in 
clusters.

\acknowledgments We thank Drs. Alexei Kritsuk and Brian O'Shea for many 
useful discussions, and the referee for useful comments. Work at Tel 
Aviv University was supported by ISF grant 225/03.

\end{document}